\documentclass[final,3p,times,twocolumn]{elsarticle}
\usepackage{graphics}
\usepackage{epsfig}

\usepackage{amssymb}
\usepackage{lineno}

\usepackage{subfigure}

\journal{Nuclear Instruments and Methods A}

\begin{document}

\begin{frontmatter}



\title{The ATLAS Pixel Insertable B-Layer (IBL)}
\author{F. H\"ugging\corref{cor1}\fnref{fh}}
\ead{huegging@physik.uni-bonn.de.}
\cortext[cor1]{On behalf of the ATLAS Collaboration}
%\fntext[fh]{On leave from elsewhere}
\address{Physikalisches Institut,  Universit\"at Bonn, Nu\ss allee 12, D-53115 Bonn}


\begin{abstract}
The ATLAS Detector will be upgraded for higher intensity running of the LHC. A long shutdown is envisioned in 2016 prior to the so-called Phase I running. A new pixel layer, called the Insertable B-Layer (IBL), will be inserted at a radius of about 3.2~cm between the existing Pixel Detector and a new (smaller radius) beam-pipe. The IBL requires the development of several new technologies to cope with the increased radiation level and pixel occupancy, as well as to improve the physics performance of the existing Pixel Detector. The IBL project provides a test of technologies for the Phase II upgrade of the entire ATLAS tracker for luminosities around $10^{35}~$cm$^{-2}$s$^{-1}$. An overview of the project with particular emphasis on the IBL layout and expected performance as well as the module development including hybridization technologies is presented.
\end{abstract}


\end{frontmatter}

%\linenumbers

\section{Introduction}
\label{sec:intro}
\begin{figure*}[ht]
\centering
\subfigure[]{
\includegraphics[width=0.91\columnwidth]{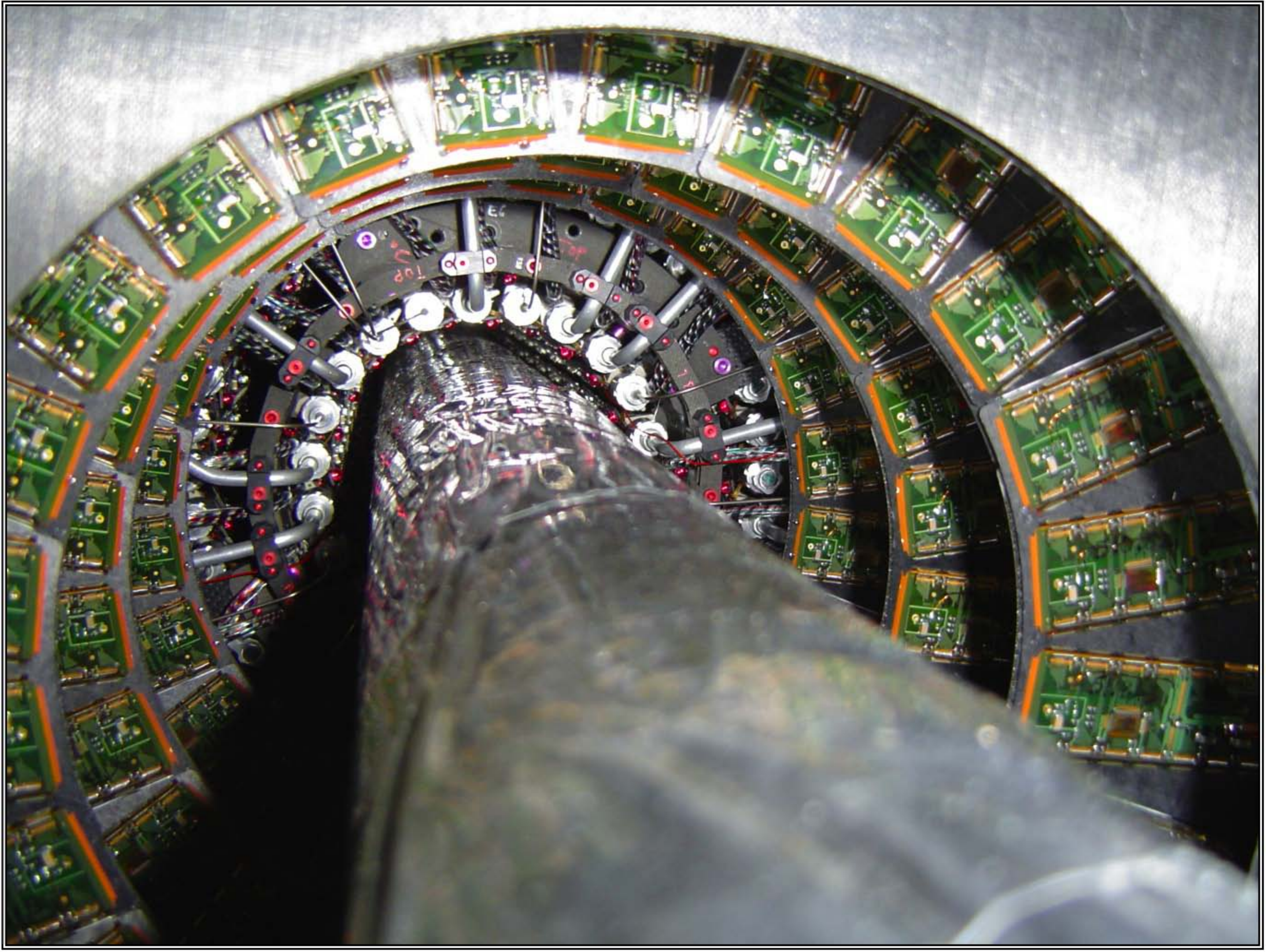}
\label{fig:pix_det}
}
\subfigure[]{
\includegraphics[width=0.95\columnwidth]{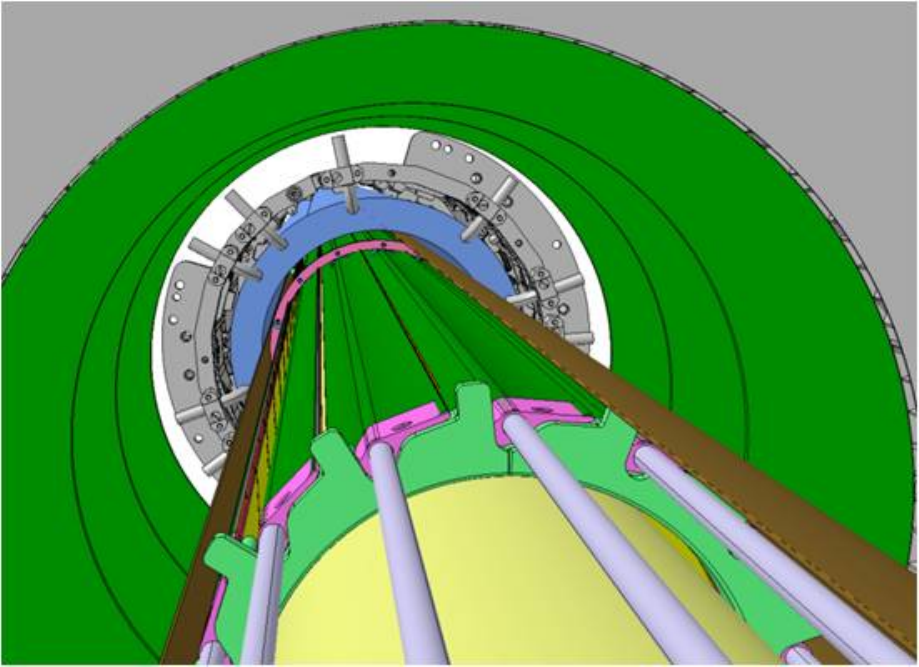}
\label{fig:ibl_det}
}
\label{fig:ibl_pix_det}
\caption{(a) Photo of the Pixel Detector with the inserted beam pipe during the integration of the present detector,
and (b) rendering of the insertion of the IBL with the smaller beam pipe~\cite{IBL-TDR:2010}.}
\end{figure*}

The ATLAS Pixel Detector~\cite{ATLAS:2008,ATLPix:2008} is the innermost detector component of the ATLAS tracking system consisting of 3 barrel layers and 6 disk layers, 3 disks in each of the forward and backward directions. It provides at least three space point measurements per track with high accuracy as needed for track and vertex determination. The layer closest to the beam pipe, the B-layer, is crucial for tracking, vertexing, and b-tagging capabilities of ATLAS especially at high luminosity. Due to the harsh radiation environment induced by the LHC the performance of the detector will deteriorate with time. The B-layer will be the first one degrading in terms of efficiency with increasing radiation damage. The expected lifetime of the B-layer is about 300~fb$^{-1}$ or 5 years. In order to keep the performance of the Pixel Detector with increasing luminosity an upgrade of the innermost pixel layer is foreseen for the long shutdown of LHC during 2016 (Phase I upgrade). The time needed to replace the existing B-layer is more than one year because of the long cooling down time of the activated material inside the detector. It was therefore decided to introduce a fourth pixel layer inside the existing detector. As one can see in Fig.~\ref{fig:ibl_pix_det}, this new B-layer can only fit into the present Pixel Detector thanks to a new smaller radius beam-pipe. The new pixel layer provides an additional space point very close to the interaction point which keeps the performance of the tracking when the B-layer starts to degrade.

This new pixel layer, called Insertable B-Layer (IBL), is currently under development. The main challenges for the development are the increased radiation damage and higher particle density closer to the interaction point at increased luminosity after the Phase I upgrade of LHC. In order to keep or even improve the performance of the Pixel Detector, several changes to the design of the hybrid pixel system are envisaged: the pixel size is reduced, the material budget is minimized by using new lightweight mechanical support materials and a CO$_2$ based cooling system is needed. The main component of the module development for the IBL is the new ATLAS pixel readout chip, FE-I4~\cite{FE_I4:2010}, designed in 130~nm technology. The new readout chip features an array of $80 \times 336$ pixels with a pixel size of $50 \times 250~\mu$m$^2$. Because of the increased pixel occupancy, the digital readout architecture has been completely redesigned. A scheme with a 4 pixel digital region has been introduced which stores the hit information until the L1 trigger decision is taken.

For the pixel sensors three different promising sensor technologies are currently under investigation. These are: planar n-in-n or n-in-p silicon sensors, full 3D silicon sensors with active edges, and pixel sensors made of polycrystalline CVD diamonds. All sensor candidates promise high radiation tolerance together with a minimal inactive area to allow efficient module placement. The challenge for the mechanics is the construction of lightweight and robust support and cooling structures which fit in the limited space between the beam pipe and the present Pixel Detector. They need to be inserted with high precision into the complex environment of the ATLAS Inner Detector. In the end the IBL will be a part of the existing Pixel Detector and therefore must be compatible with the existing off-detector readout, control and operation system. This paper outlines the status of the design of the IBL and describes the most important components of the module developments.

\section{Layout}
\label{sec:layout}
\begin{figure}[ht]
  \begin{center}
  \resizebox{7.5cm}{!}{\includegraphics{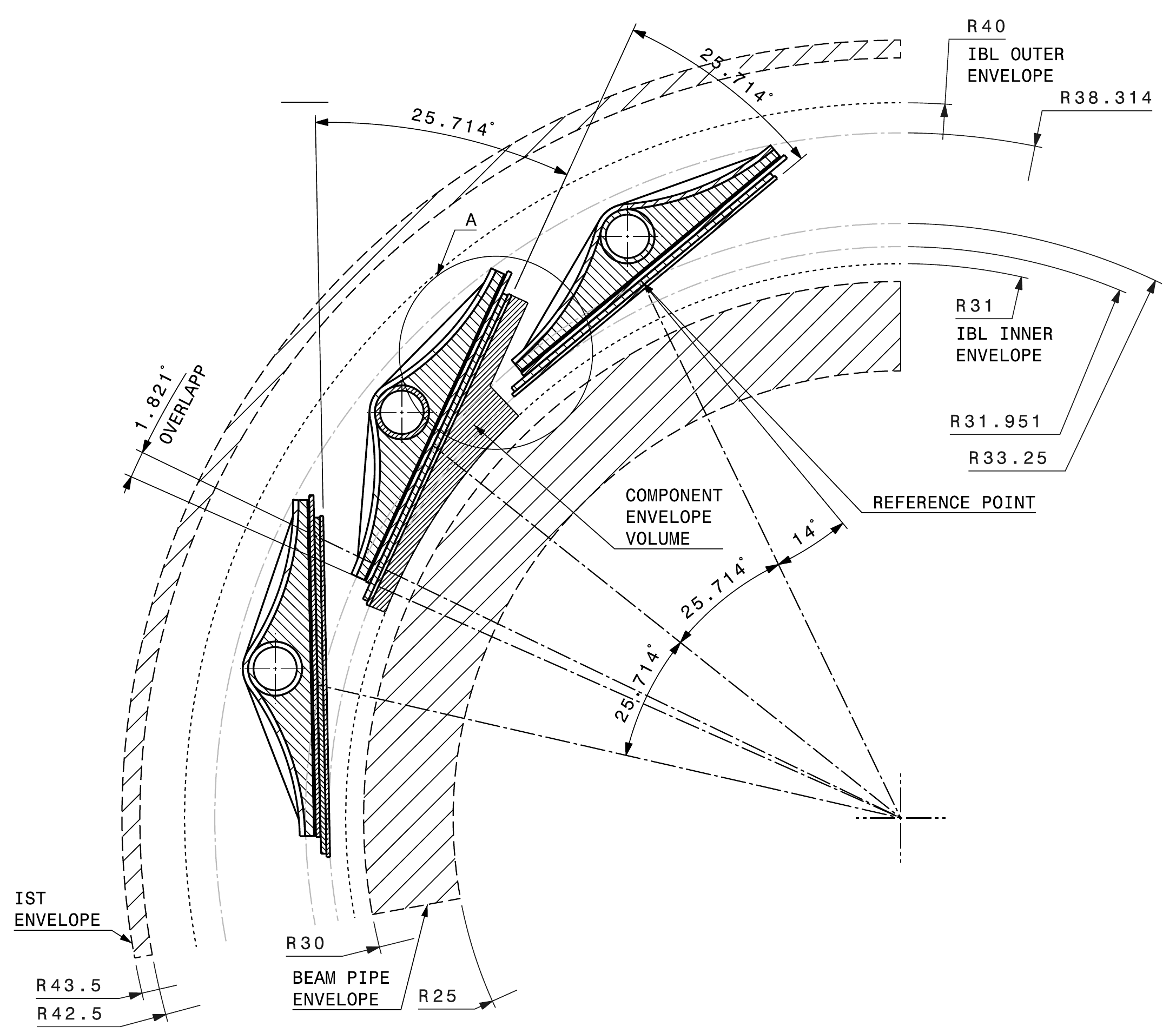}}
  \caption{Layout of the Insertable B-Layer in r$\varphi$ view.}
  \label{fig:ibl_layout}
  \end{center}
\end{figure}

The baseline layout of the IBL is a barrel layer consisting of 14 staves; a section of  the IBL is shown in Fig.~\ref{fig:ibl_layout}. The average radial distance of the sensitive area from the beam pipe is 33~mm whereas the total envelope of the IBL in radius is between 31 and 40~mm. Each stave is equipped with 16 or 32 modules depending on the sensor size. In the first case, each module will consist of 2 front-end chips on one common sensor tile, while in the second case each module will be made of one front-end chip attached to a single sensor tile. The staves are tilted by 14$^{\circ}$ to ensure a fully hermitic coverage in $\varphi$ for high p$_T$ tracks. However, due to the small radius of the IBL the sensors are at an angle between 0 and 27${^\circ}$ with respect to the radial direction.

The present Pixel Detector reaches a full geometrical coverage in z by tilting the modules along the beam axis and partially overlapping them. This cannot be done for the IBL because of space constraints. The solution chosen is to minimize the gap between modules by using a sensor design with slim or active edges. Finally, the material budget is reduced as much as possible. The goal is to reach about 1.5\% of X$_0$ for the IBL which is only slightly more than half of the X$_0$ per layer of the existing Pixel Detector B-layer. Table~\ref{tab:ibl_layout} summarizes the main layout parameters.

\begin{table*}[ht]
  \begin{center}
        \caption{Main IBL layout parameters, data taken from~\cite{IBL-TDR:2010}.}
        \label{tab:ibl_layout}
%        \vspace{0.1in}
        \begin{tabular}{l c c}
        \hline
                                                                            &Value                  &Unit\\
        \hline
                Number of staves                                            &14                     &\\
                Number of modules per stave (single/double FE-I4)           &32 / 16                &\\
                Pixel size($\varphi$, z)                                    &50, 250                &$\mu$m\\
                Module active size W$\cdot$L (single/double FE-I4)          &16.8$\times$40.8 /16.8$\times$20.4 &mm$^2$\\
                Coverage in $\eta$, no vertex spread                        &$|\eta| < 3.0$         &\\
                Coverage in $\eta$, 2$\sigma$ (=112 mm) vertex spread       &$|\eta| < 2.58$        &\\
                Active z extent                                             &330.15                 &mm\\
                Geometrical acceptance in z (min, max)                      &97.4, 98.8             &\%\\
                Stave tilt angle in $\varphi$ (center of sensor, min, max)  &14.00, -0.23, 27.77    &degree\\
                Overlap in $\varphi$                                        &1.81                   &degree\\
                Sensor thickness (depending on technology)                   &150 - 600           &$\mu$m\\
                Radiation length at z = 0                                   &1.54                   &\% of X$_0$\\
        \hline
        \end{tabular}
  \end{center}
\end{table*}

\section{Performance}
\label{sec:perf}
The physics impact of the IBL has been studied in detail by fully integrating the IBL into the ATLAS Inner Detector simulation software~\cite{IBL-TDR:2010}. This allows Monte Carlo performance studies and comparisons to be done with or without the IBL integrated to the current Pixel Detector. Because of the low mass and close proximity to the interaction point, the IBL improves the quality of the impact parameter reconstruction for tracks and thereby the vertexing and $b$-tagging performance. For example, the rejection of light jets in $t\bar{t}$ events without pileup for a $b$-tagging efficiency of 60\% shows an increase close to 2 with the IBL installed, as can be seen in table~\ref{tab:b_tagging}. The jet rejection is defined as the inverse of the mis-tagging rate which is the fraction of non $b$-jets that are tagged as $b$-jets~\cite{ATLAS-Perf:2009}.

\begin{table}[ht]
  \begin{center}
        \caption{Rejection of light jets in $t\bar{t}$ events without pileup for a $b$-tagging efficiency of 60\%, data taken from~\cite{IBL-TDR:2010}. The IP3D algorithm combines transverse and longitudinal impact parameter information. The IP3D+SV1 algorithm associates the IP3D with a vertex tagging algorithm based on an inclusive secondary vertex search~\cite{ATLAS-Perf:2009}.}
        \label{tab:b_tagging}
%        \vspace{0.1in}
        \begin{tabular}{l c c c}
        \hline
                Algorithm       &Without IBL     &With IBL       &Ratio \\
        \hline
                IP3D           &83$\pm$1.5     &147$\pm$3.4   &1.8 \\
                IP3D + SV1     &339$\pm$12     &655$\pm$32    &1.9 \\
        \hline
        \end{tabular}
  \end{center}
\end{table}

The studies also showed that with the IBL the track and vertex reconstruction is robust against pileup and hard failures of modules in the existing B-layer or in other silicon layers. The $b$-tagging performance of ATLAS with the IBL at Phase I luminosity pileup is comparable to the current detector performance without pileup. The impact parameter resolution is recovered for all studied scenarios with detector defects. The IBL also improves the $b$-tagging performance even in case of a B-layer failure. In summary, the IBL will lead to an improved sensitivity of ATLAS during Phase I for signals in physics channels which involve $b$-jets such as for a low mass Standard Model Higgs in the channel $H\rightarrow b\bar{b}$.

\section{Modules}
\label{sec:modules}
The basic unit of the IBL is a module which consists of one or two front end chips bump bonded to one sensor. This module is a single entity from the point of view of mechanics, DAQ, detector control system and services. However, the IBL module physical size depends on the still to be chosen sensor technology. If planar or diamond sensors are chosen, the module will consist of two front end chips and one sensor (2-chip module). If 3D sensors are used, the module will be made of one chip and one sensor (1-chip module). The I/O, services, and control modularities are independent of sensor technology. The DAQ unit always consists of two front end chips, with common clock and control inputs, and two data outputs. It should be noted that this configuration matches the mechanical module in case of planar or diamond sensors, but not in the case of 3D sensors. The IBL power and sensor bias service units consist of four front end chips in parallel.

\begin{figure*}[ht]
\centering
\subfigure[]{
\includegraphics[width=\columnwidth]{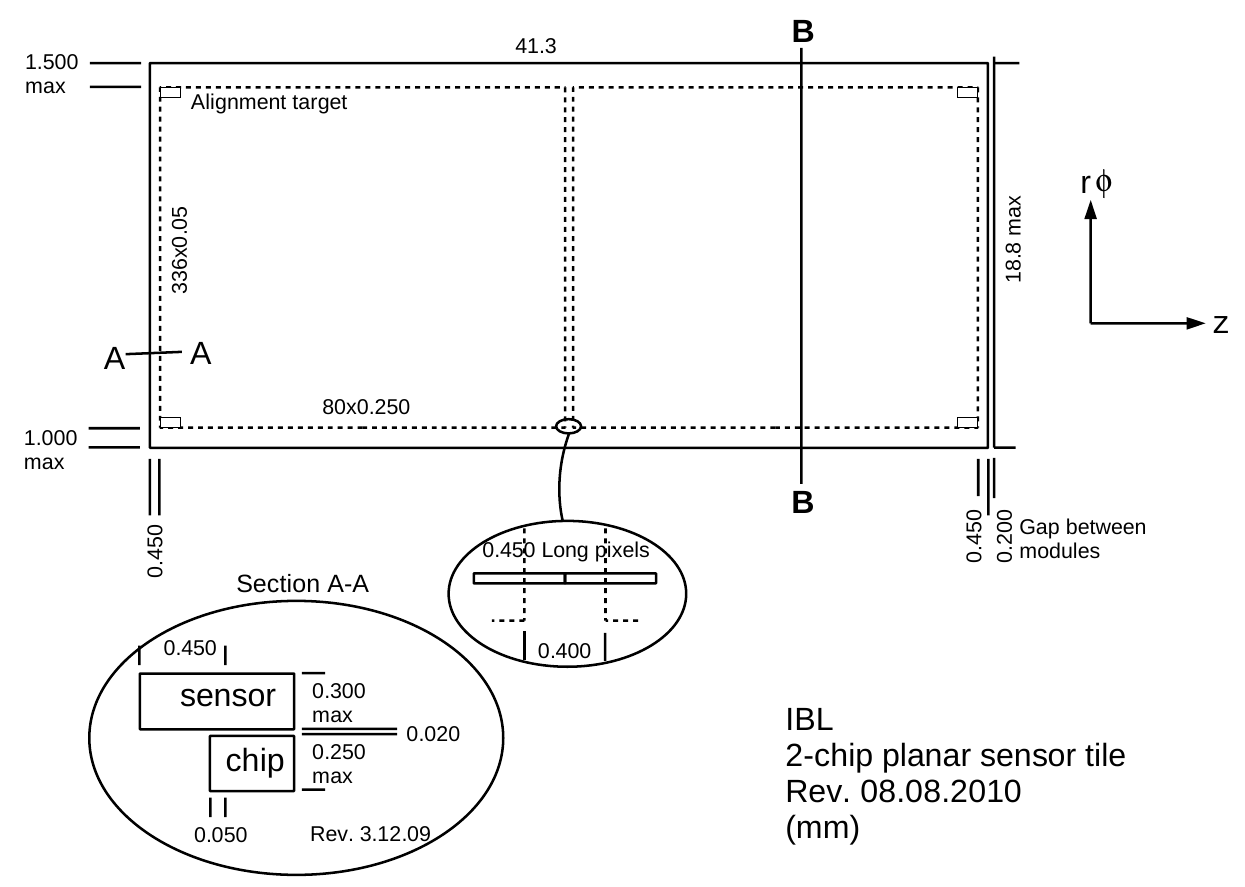}
\label{fig:2_chip}
}
\subfigure[]{
\includegraphics[width=0.9\columnwidth]{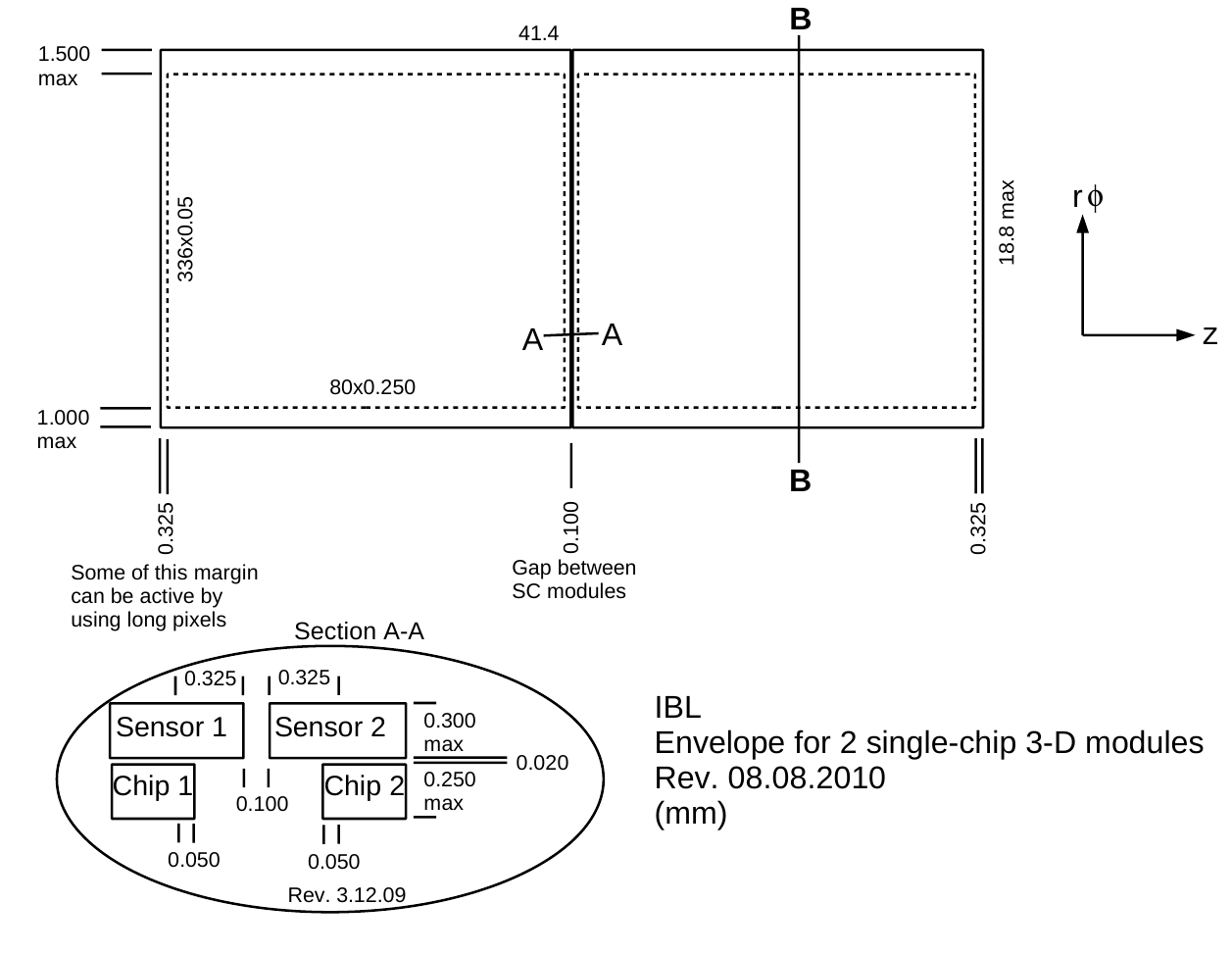}
\label{fig:singlechip}
}
\label{fig:mod_concepts}
\caption{Module format of (a) a 2-chip module, and (b) 2 1-chip modules. The dotted lines indicate the size of the active area covered by one front-end chip but the outline of the front-end chips is not drawn. The sections labeled B-B are identical and not shown.}
\end{figure*}

The IBL module outline using 2-chip assemblies is shown in Fig.~\ref{fig:2_chip}. If 3D sensors are chosen, a similar outline can be produced using 1-chip assemblies with narrow sensor edges as shown in Fig.~\ref{fig:singlechip}. Active edge technology to produce such narrow edges is a feature of 3D sensors. There is no shingling in z of the IBL modules due to the extreme radial space constraints, and this results in slightly different geometric acceptance for 2-chip and 1-chip assemblies. For 1-chip (2-chip) assemblies the nominal acceptance for particles normal to the beam is 98.8\% (97.4\%). An air gap of 100~$\mu$m (200~$\mu$m) between 1-chip (2-chip) assemblies has been assumed to take into account the higher bias voltages needed by 2-chip (planar or diamond) sensors with respect to 3D assemblies. The 100~$\mu$m gap is considered adequate for 150~V (in a dry atmosphere) and the 200~$\mu$m gap is considered sufficient for placement of polyimide insulation film if necessary. But both methods still require validation with prototypes.

The module assembly starts by bump bonding the sensor to the front end chip. The module is then dressed by adding a flexible printed circuit board (flex) which acts as an electrical interface to the outer world. No active electrical components are used for the upgraded module flex unlike in the present ATLAS system where a module control chip steers several front end chips.

\subsection{Sensors}
\label{sec:sensors}
As already mentioned, there are three candidate sensor technologies that may meet the IBL requirements: either n-in-n or n-in-p planar silicon sensors, 3D silicon sensors, and diamond sensors. For all of these candidate sensor technologies detailed qualification studies are under way. The status of these investigations as well as the specific description of the sensor types can be found elsewhere~\cite{Planar_n_in_n:2010,Planar_n_in_p:2010,3D:2010,diamant:2010}.
%~\cite{Planar_n_in_n:2010}~\cite{Planar_n_in_p:2010}~\cite{3D:2010}~\cite{diamant:2010}

Planar sensors as used for the ATLAS tracker feature very well understood mechanical properties, reliable manufacturing sources, and are available at relative low cost with high production yield. For IBL either n-in-n sensors with standard thickness of 250~$\mu$m, or 150~$\mu$m thin n-in-p sensors are considered. The n-in-n sensors are similar to the sensor used in the present ATLAS Pixel Detector. It is expected that they can withstand the higher radiation damage of $5\cdot 10^{15}$~n$_{\mathrm{eq}}$cm$^{-2}$ for IBL. The thin n-in-p sensors take advantage of the increased radiation tolerance of thin p-type silicon as they feature a higher electrical field strength at a given maximum bias voltage. Hence both planar sensor types require the lowest operating temperature and highest bias voltage.

3D sensors are sufficiently radiation tolerant because of the reduced charge collection distance between the 3D electrodes (about $70~\mu$m for the IBL sensors). Therefore, they require only moderate maximum bias voltages and intermediate operating temperature. But the production of 3D sensors with high yield and good uniformity is still to be demonstrated.

Diamond sensors offer the required radiation tolerance with the least amount of cooling needed due to the absence of leakage current after irradiation. But they require a similar maximum bias voltage as the planar sensor and the production at moderate cost with high yield and good uniformity must still be proven.

\begin{table}[ht]
  \begin{center}
        \caption{Thickness and inactive edge area for the sensors under consideration for IBL.}
        \label{tab:sensor_parameter}
%        \vspace{0.1in}
        \begin{tabular}{l c c}
        \hline
                Sensor type             &Thickness [$\mu$m] &Edge width [$\mu$m]\\
        \hline
                Planar n-in-n           &250                &100 - 450\\
                Planar n-in-p           &150                &100 - 450\\
                3D single sided         &230$\pm$15        &50\\
                3D double sided         &230$\pm$15        &200\\
                pCVD diamond            &400 - 600       &100 - 200\\
        \hline
        \end{tabular}
  \end{center}
\end{table}

The sensor thickness and the inactive edge area varies for each sensor type and thus has an impact on the design of the support structures. In table~\ref{tab:sensor_parameter}, these parameters are summarized for all sensor types. Planar sensors can be made thin enough, but require the widest edges so far (450~$\mu$m). There are therefore developments pursued in the planar sensor community to shrink these edges to below 200~$\mu$m, which is desirable for the usage in IBL. 3D sensors already feature narrow edge design for single sided and double sided processing and are relatively thin. Diamond detectors can also be made with narrow edges but suffer from the relative big thickness (at least 400~$\mu$m) needed for sufficient charge collection. This bigger thickness sets higher requirements on the geometrical clearance of the IBL but is still acceptable in terms of radiation length.

The charge collection with increasing fluence is one of the most important parameters to measure the sensor performance for each sensor technology. But the sensors's charge collection is not the decisive parameter by itself. The combined electrical performance of the FE-I4 chip with the sensor may influence the required charge at the sensor's end of lifetime. For instance, the minimum stable operating threshold of the FE-I4 may increase the required charge after irradiation. Therefore all investigation of the three sensor types must be eventually done with devices bump bonded to FE-I4.

\subsection{Electronics}
\label{sec:electronics}
The IBL module format is based on a new integrated circuit (FE-I4). The front end chip used in the present detector (FE-I3~\cite{ATLPix:2008}) was excluded from IBL use due to two fundamental problems: the hit rate capability and radiation hardness are not high enough, and the active fraction of the footprint is too small to build a compact layer with high geometric acceptance. FE-I4 was designed to address these problems as well as to make progress towards lower cost pixel detectors needed for an eventual replacement of the complete ID for sLHC. The close to 90\% active footprint of FE-I4 (compared to 74\% for FE-I3) enables the design of a low radial profile layer, as required for IBL. Table~\ref{tab:I3-I4} summarizes the main differences between the FE-I3 and the FE-I4 readout chip.

%\begin{figure}[ht]
%  \begin{center}
%  \resizebox{7.5cm}{!}{\includegraphics{bilder/fig_fe_i4.pdf}}
%  \caption{FE-I4 chip outline.}
%  \label{fig:fe_i4}
%  \end{center}
%\end{figure}

\begin{table}[ht]
  \begin{center}
        \caption{Main differences between the FE-I3 and FE-I4 readout chip.}
        \label{tab:I3-I4}
%        \vspace{0.1in}
        \begin{tabular}{l c c c}
        \hline
                                    &FE-I3          &FE-I4              &Unit \\
        \hline
                Pixel size          &50$\times$400         &50$\times$250             &$\mu$m$^2$\\
                Pixel array         &18$\times$160         &80$\times$336             &\\
                Chip size           &7.6$\times$10.8       &20.2$\times$19.0          &mm$^2$\\
                Active fraction     &74             &89                 &\%\\
                Analog current      &26             &10                 &$\mu$A/pix.\\
                Digital current     &17             &10                 &$\mu$A/pix.\\
                Analog voltage      &1.6            &1.4                &V\\
                Digital voltage     &2.0            &1.2                &V\\
                Pseudo-LVDS speed   &40             &160                &Mb/s\\
        \hline
        \end{tabular}
  \end{center}
\end{table}

Additionally, the FE-I4 is a self contained electrical unit requiring no module control chip to attain the complete module functionality. Within the IBL module 2 chips are controlled in parallel (shared clock and command inputs), but each one has a dedicated data output, leading to one clock input, one command input, and 2 data outputs. In terms of number of signals, this is the same as for a present detector B-layer module. Furthermore, the command and clock inputs are fully compatible with the present detector protocols, even if command bitcodes are slightly different. It is therefore possible to control an IBL module using present detector hardware (with software and firmware changes). The IBL module data outputs, on the other hand, must have a higher bandwidth than those used in the present B-layer to handle the increased hit rate. The output links have therefore been designed with a new protocol that is not fully compatible with the existing detector DAQ hardware. However, it is still compatible with the optical hardware used in the present detector (VCSEL, pin array and related control and driver chips).

More detailed information about the design and the submission of the first full scale prototype of the FE-I4 can be found in~\cite{FE_I4:2010,IBL-TDR:2010}. One of the main purposes of the full scale prototypes will be the building of bump bonded devices with each sensor type. Thus an evaluation of IBL like modules before and after irradiation will be possible which is needed for the sensor choice.

\subsection{Hybridization}
\label{sec:hybridization}
Sensors and front end chips are integrated through flip chip bump bonding. The requirements for this process which are a bump pitch of 50~$\mu$m together with a defect rate of less than 10$^{-4}$ are similar to those met with the present detector. However, the large size of the FE-I4 makes the IBL flip chip process more challenging. For the present detector, the FE-I3 chips were thinned to 190~$\mu$m which was thick enough to control the bow of the chips within the bump height tolerance when changing between room temperature and solder reflow temperature (250$^{\circ}$C). Because of the larger size, the FE-I4 chip would have to be approximately 400~$\mu$m thick to achieve the same bow control~\cite{Low_material:2010}. Since this is unacceptable for the IBL, a modified flip chip process must be used. The modification can entail the use of a support wafer temporarily bonded to the thinned FE-I4 chips in order to control bow during reflow, and which must be removed before flex module assembly.

\begin{figure}[ht]
  \begin{center}
  \resizebox{7.5cm}{!}{\includegraphics{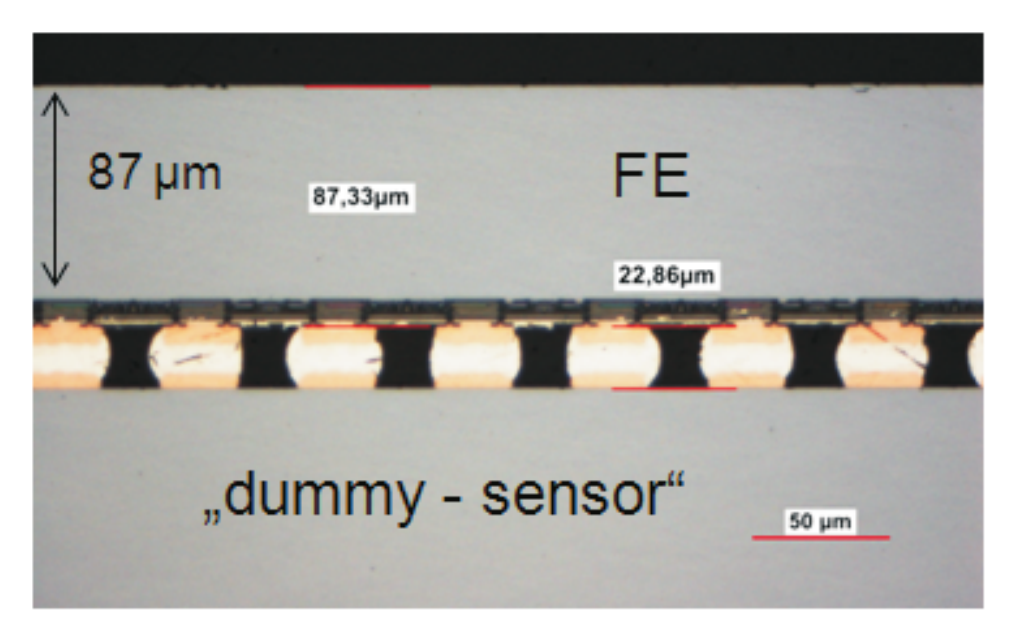}}
  \caption{Cross-section of a flip chip test assembly of FE-I4 size, thinned down to $87~\mu$m and bump bonded
using a glass carrier wafer as described in the text. The $87~\mu$m includes the dark area just above the
bumps, which is the IC circuitry. The gap between chip and sensor (height of the connected bumps) is shown
as $22.86~\mu$m.}
  \label{fig:thin_mod}
  \end{center}
\end{figure}

This technique has been demonstrated on FE-I4 sized wafer fragments of FE-I3 wafers. Such samples have been thinned to about 90~$\mu$m and bump bonded to dummy sensors with high bump yield. The samples were made of FE-I3 chips cut in 2-by-2 arrays (roughly $15 \times 22$~mm$^2$). After the solder bump deposition, the FE-I3 IC wafer was thinned using standard wafer grinding methods and mounted onto a thick glass support wafer using polyimide film. The full assembly of glass plus IC wafer was fully diced into 2-by-2 chip arrays. The polyimide bond can withstand solder reflow temperature without losing adhesion. The IC-glass sandwich arrays were flipped onto the dummy sensors without any thermally induced bow. After flip-chip, the glass support was removed by laser exposure of the polyimide film through the glass. Fig.~\ref{fig:thin_mod} shows a cross-section picture of a flip-chipped device after support wafer removal. One can clearly recognize the uniform ball shaped solder bump bonds with a pitch of 50~$\mu$m. No evidence for disconnected bumps has been found in the full assembly. The effective readout IC thickness is measured to be 87~$\mu$m. The handling of these devices is demanding and subsequent assembly steps like gluing to the support structure and wire bonding can easily lead to disconnected bump bonds.

\begin{figure}[ht]
  \begin{center}
  \resizebox{7.5cm}{!}{\includegraphics{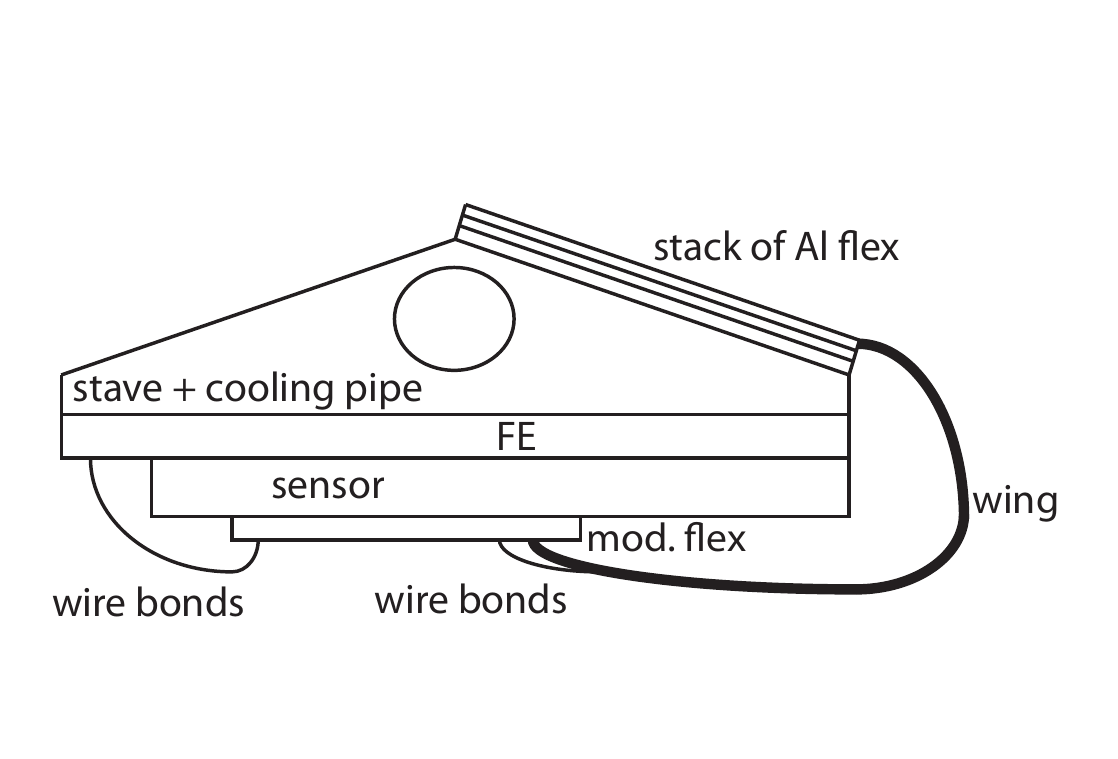}}
  \caption{Stave flex to module attachment via a kapton wing and wire bonds.}
  \label{fig:mod_on_stave}
  \end{center}
\end{figure}

This bare module is made into a fully stand-alone detector unit by adding a flex with passive components, a method of external connection, and a support frame. The construction of a stand-alone object in a disposable support frame was extremely useful for testing, burn-in, and general handling during the module production for the present detector, and so this procedure should be maintained for IBL. The important development needed for IBL is the elimination of individual module connectors. An IBL stave will have 32 (16) single chip (2-chip) modules. A pigtail style connector on each one (as used in the present detector barrel) would translate into a large amount of dead material and would require a significant amount of space which does not exist. Conversely, an individual, long cable permanently attached to each
module leading to a distant connector would bring serious complications for assembly and would require challenging strain relief in the active region.

The solution under development for the IBL is therefore to have cables already integrated on the stave mechanical structure before any modules are loaded, and to connect such cables to each module by some permanent, reworkable method, such as wire bonding as shown in Fig.~\ref{fig:mod_on_stave}. The module flex would be very small with bond pads within the sensor perimeter or on a very short tail. A small flex mounted inside a handling frame can cover less than half of a single chip bare module. The final flex module is cut away from that PCB handling frame after all tests and burn-in have been completed. This module design is compatible with a stave with integrated cable featuring flex "flap" at the position of each module, to be affixed to the bare part of the sensor and then wire bonded to the flex~\cite{IBL-TDR:2010}.

%\section{Mechanics}
%\label{sec:mech}
%
%\begin{figure*}[ht]
%\centering
%\subfigure[]{
%\includegraphics[width=0.95\columnwidth]{bilder/fig_stave_x_sec.pdf}
%\label{fig:stave_x_sec}
%}
%\subfigure[]{
%\includegraphics[width=0.95\columnwidth]{bilder/fig_stave_3d.pdf}
%\label{fig:stave_3d}
%}
%\label{fig:mod_concepts}
%\caption{(a) Cross section and (b) 3D model of the IBL stave~\cite{IBL-TDR:2010}. Dimensions are in millimeters.}
%\end{figure*}

\section{Conclusions}
\label{sec:conclusion}
It is foreseen to install an additional, innermost pixel layer (IBL) to the present Pixel Detector in ATLAS to cope with the increased luminosity during Phase I upgrade of LHC. This new pixel layer will improve the impact parameter resolution and will as well be able to compensate any degradation of the current Pixel Detector B-layer due to radiation damage, ageing or increasing data rates. The R\&D for the IBL project is well advanced and a TDR has been recently published~\cite{IBL-TDR:2010}. New sensor developments are evaluated, a new front end chip (FE-I4) has been submitted and an appropriate, flexible module design is pursued. The integration of this new pixel layer includes new lightweight support structures and low material interconnection techniques. The installation of the IBL is currently scheduled for the long LHC shutdown in 2016.



%% References


\bibliographystyle{elsarticle-num}
\bibliography{mybib}






% \end{thebibliography}

\end{document}